\documentstyle[aps,prl,twocolumn,epsfig]{revtex}
\begin{document}

\title{Neutral Color Superconductivity and Pseudo-Goldstone Modes}
\author{Lianyi He, Meng Jin and Pengfei Zhuang\\
        Physics Department, Tsinghua University, Beijing 100084, China}
\maketitle

\begin{abstract}
Four of the five expected
Goldstone modes, which will be eaten up by gauge fields, in
neutral two-flavor color superconductor are actually
pseudo-Goldstone modes, and their degenerated mass is exactly the
magnitude of the color chemical potential, which is introduced to
guarantee the color neutrality at moderate baryon density.
\end{abstract}
\\

Since the attractive interaction in the antitriplet quark-quark
channel in quantum chromodynamics (QCD), the cold and dense quark
matter is believed to favor the formation of diquark condensate
and in the superconducting phase\cite{cscreview}. In the idealized
case at asymptotically high baryon density, the color
superconductivity with two massless flavors and the
color-flavor-locking (CFL) phase with three degenerated massless
quarks have been widely discussed from first principle QCD
calculations\cite{highu}. For physical applications, one is more
interested in the moderate baryon density region which may be
realized in compact stars and, in very optimistic cases, even in
heavy-ion collisions. To have a stable and macroscopic color
superconductor, one should take into account the electric and
color charge neutrality condition\cite{neutral-1,neutral-2} which
lead to a new phase, the gapless color
superconductivity\cite{gapless} or the breached pairing
phase\cite{bp}. The most probable temperature for this new phase
is finite but not zero\cite{calabash}. In two flavor case, the
color neutrality can be satisfied by introducing a color chemical
potential $\mu_8$ in the four-fermion interaction theory at
moderate baryon density\cite{neutral-2}, or by a dynamic
generation of a condensation of gluon field $A_0^8$ in the frame
of perturbative QCD at extremely high baryon
density\cite{neutral-3} where the back ground gluon field $\langle
A_0^8\rangle$ plays the role of the color chemical potential
$\mu_8$.

It is generally accepted that in the two-flavor color
superconductor, there will be five massless Goldstone bosons,
corresponding to the spontaneously broken color symmetry from
$SU_C(3)$ to $SU_C(2)$. At moderate baryon density, if the charge
neutrality condition is not taken into account, there are only
three Goldstone modes\cite{goldstone-1}. Since the degenerated
mass of two diquarks of the rest five collective modes is
proportional to the net color charge $Q_8$ of the system, it is
expected that\cite{goldstone-1} one can recover the five massless
Goldstone bosons by requiring color neutrality. While the
Goldstone bosons will finally be eaten up by gauge fields through
Higgs mechanism\cite{goldstone}, it is necessary to check whether
the five expected Goldstone modes are really massless. In this
Letter we will show that four of the five expected Goldstone modes
are actually pseudo-Goldstone modes, their degenerated mass is
exactly the magnitude of the color chemical potential $\mu_8$
which is used to guarantee the color neutrality of the system.

The pseudo-Goldstone bosons were generally discussed in theories
with spontaneously broken local symmetries thirty years
ago\cite{weinberg}. The higher-order correction leads to spinless
bosons which behavior like Goldstone bosons but have a small mass.

For a neutral two-flavor color superconductor, the quark chemical
potential matrix
\begin{eqnarray}
\label{mu1}
\mu=diag(\mu_{u1},\mu_{u2},\mu_{u3},\mu_{d1},\mu_{d2},\mu_{d3})
\end{eqnarray}
in color and flavor space can be expressed in terms of baryon
chemical potential $\mu_b$, electrical chemical potential $\mu_e$,
and color chemical potential $\mu_8$,
\begin{eqnarray}
\label{mu2}
&&\mu_{u1}=\mu_{u2}=\mu_b/3-2\mu_e/3+\mu_8/3\ ,\nonumber\\
&&\mu_{u3}=\mu_b/3-2\mu_e/3-2\mu_8/3\ ,\nonumber\\
&&\mu_{d1}=\mu_{d2}=\mu_b/3+\mu_e/3+\mu_8/3\ ,\nonumber\\
&&\mu_{d3}=\mu_b/3+\mu_e/3-2\mu_8/3\ ,
\end{eqnarray}
where $\mu_b$ controls the baryon number density, and $\mu_e$ and
$\mu_8$ have to be introduced to ensure the electric and color
charge neutrality. If the $SU_C(3)$ color symmetry is not broken,
the color chemical potential $\mu_8$ has to vanish, otherwise it
would break $SU_C(3)$ explicitly. However, when the color symmetry
is broken spontaneously by a color-charged diquark condensate,
$\mu_8$ does not need to be zero\cite{rischke}.

It is well-known that a nonzero quark-antiquark condensate
$\langle\bar\psi\psi\rangle$ spontaneously breaks the chiral
symmetry $SU_L(2)\bigotimes SU_R(2)$ of the system, and the
corresponding Goldstone bosons are the three pion mesons. However,
when the current quark mass $m\neq 0$, the pions get a small mass
proportional to $m$, due to the explicit chiral symmetry breaking.
Similarly, when $\mu_8=0$, a nonzero diquark condensate $\langle
\bar\psi^c_{i\alpha}i\gamma^5\epsilon^{ij}\epsilon^{\alpha\beta
3}\psi_{j\beta}\rangle$ spontaneously breaks down the color
symmetry $SU_C(3)$ to $SU_C(2)$, and causes five Goldstone bosons
corresponding to the broken generators $T_4,T_5,T_6,T_7$ and
$T_8$, where $\epsilon^{ij}$ and $\epsilon^{\alpha\beta\gamma}$
are totally antisymmetric tensors in flavor and color space,
respectively, and it is assumed that only the first two colors
participate in the condensate, while the third one does not.
However, in presence of a nonzero $\mu_8$, the $SU_C(3)$ symmetry
is explicitly broken down to $SU_C(2)\bigotimes U_C(1)$ with
broken generators $T_4,T_5,T_6$ and $T_7$. Therefore, the four
Goldstone bosons corresponding to the broken generators
$T_4,T_5,T_6$ and $T_7$ will become massive, and only the
Goldstone boson corresponding to the broken generator $T_8$
remains massless.

Here we have not considered the color chemical potential $\mu_3$,
since the neutrality for the color charge $Q_3$ is automatically
satisfied in the current case with diquark condensation in the
third color direction.

To investigate quantitatively the diquarks as pseudo-Goldstone
bosons, we have to choose a suitable model in describing color
superconductivity at moderate baryon densities. It is generally
accepted that the Nambu--Jona-lasinio model (NJL)\cite{njl}
applied to quarks\cite{njlreview} offers a simple but effective
scheme to study chiral symmetry restoration\cite{hufner}, color
symmetry spontaneously
breaking\cite{neutral-2,gapless,calabash,goldstone-1,schwarz,ruster}
and isospin symmetry spontaneously
breaking\cite{barducci,he-1,he-2}. In the mean field approximation
to quarks and random phase approximation (RPA) to mesons, one can
obtain the hadronic mass spectra and the static properties of
mesons remarkably well\cite{njlreview}, especially the Goldstone
modes corresponding to the chiral symmetry spontaneously
breaking\cite{hufner} and to the isospin symmetry spontaneously
breaking\cite{he-2}.

The flavor $SU(2)$ NJL model is defined through the Lagrangian
density
\begin{eqnarray}
\label{njl} {\cal L} &=&
\bar{\psi}\left(i\gamma^{\mu}\partial_{\mu}+\mu\gamma_0\right)\psi
+G_S\left[\left(\bar{\psi}\psi\right)^2+\left(\bar{\psi}i\gamma_5\vec{
\tau}\psi\right)^2 \right]\nonumber\\
&+&G_D\left(\bar\psi^c_{i\alpha}
i\gamma^5\epsilon^{ij}\epsilon^{\alpha\beta
\gamma}\psi_{j\beta}\right)\left(\bar\psi_{i\alpha}
i\gamma^5\epsilon^{ij}\epsilon^{\alpha\beta\gamma}\psi^c_{j\beta}\right)
\ ,
\end{eqnarray}
where $G_S$ and $G_D$ are coupling constants in color singlet
channel and color anti-triplet channel, respectively, $\psi^c =
C\bar\psi^T$ and $\bar\psi^c = \psi^T C$ are charge-conjugate
spinors, $C = i\gamma^2\gamma^0$ is the charge conjugation matrix.

Since we focus in this Letter on the color symmetry spontaneously
breaking and the corresponding Goldstone modes, to simply the
notation, we consider in the following only the color symmetry
spontaneously breaking phase with nonzero diquark condensate
\begin{equation}
\label{condensate} \Delta = -2G_D\langle
\bar\psi^c_{i\alpha}i\gamma^5\epsilon^{ij}\epsilon^{\alpha\beta
3}\psi_{j\beta}\rangle\ ,
\end{equation}
and assume that the chiral symmetry is restored in this phase.

In the mean field approximation, the quarks behavior like
quasi-particles, and the diquark condensate is controlled by the
gap equation\cite{neutral-2}
\begin{equation}
\label{gap} 1-2G_DI_\Delta = 0\ ,
\end{equation}
with the function
\begin{equation}
\label{idelta} I_\Delta = 4\int{d^3{\bf p}\over
(2\pi)^3}\sum_{\epsilon=\pm}{1-f(E^\epsilon_+)-f(E^\epsilon_-)\over
E^\epsilon_\Delta}\ ,
\end{equation}
where the quasi-particle energies are defined as $E^\pm_\mp =
E^\pm_\Delta \mp\delta\mu, E^\pm_\Delta = \sqrt{(|{\bf
p}|\pm\bar\mu)^2+\Delta^2}$ with the two effective chemical
potentials $\bar\mu$ and $\delta\mu$ given by $\bar\mu =
\mu_b/3-\mu_e/6+\mu_8/3$, and $\delta\mu=\mu_e/2$, and
$f(x)=1/\left(e^{x/T}+1\right)$ is the Fermi-Dirac distribution
function.

To consider the color and electric charge neutralities we need to
calculate the color and electric charge densities which can also
be expressed as summations of quasi-particle
contributions\cite{neutral-2},
\begin{eqnarray}
\label{charge} Q_8 &=& \int{d^3{\bf p}\over
(2\pi)^3}\sum_{\epsilon=\pm}\epsilon\Big[{E^\epsilon_0\over
E^\epsilon_\Delta}\left(1-f(E^\epsilon_+)-f(E^\epsilon_-)\right)\nonumber\\
&+&\left(f(E_{u3}^\epsilon)+f(E_{d3}^\epsilon)\right)\Big]=0\
,\nonumber\\
Q_e &=& \int{d^3{\bf p}\over
(2\pi)^3}\sum_{\epsilon=\pm}\Big[\epsilon{E^\epsilon_0\over
E^\epsilon_\Delta}\left(1-f(E^\epsilon_+)-f(E^\epsilon_-)\right)\nonumber\\
&+&3\left(f(E^\epsilon_+)-f(E^\epsilon_-)\right)-\epsilon
\left(2f(E_{u3}^\epsilon)-f(E_{d3}^\epsilon)\right)\Big]\nonumber\\
&-&{\mu_e^3\over 2\pi^2}=0\ ,
\end{eqnarray}
with quark energies $E_0^\pm = |{\bf p}|\pm\bar\mu,
E_{u3}^\pm=|{\bf p}|\pm\mu_{u3}$ and $E_{d3}^\pm = |{\bf
p}|\pm\mu_{d3}$, where the last term of $Q_e$ is the contribution
from the electron gas. Note that the quarks with color 3 are not
involved in the diquark condensation, they behavior like free
quarks in the color superconductivity phase. The diquark
condensate $\Delta$ and color and electric chemical potentials
$\mu_8$ and $\mu_e$ as functions of temperature $T$ and baryon
chemical potential $\mu_b$ are determined self-consistently by the
above three coupled equations.

We now investigate diquark and meson properties at finite
temperature and chemical potentials. In the NJL model, the diquark
and meson modes are regarded as quantum fluctuations above the
mean field. The meson modes can be calculated in the frame of
RPA\cite{njlreview,hufner}. When the mean field quark propagator
is diagonal in color, flavor, and Nambu-Gorkov\cite{ng} space, for
instance the case with only chiral condensation, the summation of
bubbles in RPA selects its specific channel by choosing at each
stage the same proper polarization function, a meson mode which is
determined by the pole of the corresponding meson propagator is
related to its own polarization function
$\Pi_{MM}(k)$\cite{njlreview,hufner} only,
\begin{equation}
\label{meson1} 1-2G_S\Pi_{MM}(k)=0\ .
\end{equation}
However, for the quark propagator with off-diagonal elements, like
the cases of $\eta$ and $\eta'$ meson spectrum\cite{njlreview},
pion superfluidity\cite{he-2}, and color superconductivity
considered here, we must consider carefully all possible channels
in the bubble summation in RPA.

In the two-flavor NJL model there are four meson modes, the scalar
meson $\sigma$ and the three pseudoscalar mesons $\pi_+, \pi_0,
\pi_-$. In the current case considering color symmetry
spontaneously breaking in the third direction in color space,
there are six kinds of diquarks, $D_1, D_2, D_3$ and $D_{\bar 1},
D_{\bar 2}, D_{\bar 3}$ constructed, respectively, by colors 2 and
3, 1 and 3, 1 and 2, $\bar 2$ and $\bar 3$, $\bar 1$ and $\bar 3$,
and $\bar 1$ and $\bar 2$.

Since we restrict ourselves in the color symmetry spontaneously
breaking phase, the mixture among the different channels in the
bubble summation in RPA is greatly reduced. The dispersion
relations for the meson modes $\sigma$ and $\pi_0$ are determined
by their own polarization functions,
\begin{eqnarray}
\label{mass1}
&&1-2G_S\Pi_{\sigma\sigma}(k) = 0\ , \nonumber\\
&&1-2G_S\Pi_{\pi_0\pi_0}(k) = 0\ ,
\end{eqnarray}
while the ones for the other mesons $\pi_+$ and $\pi_-$ and all
the diquarks satisfy the coupled equations,
\begin{eqnarray}
\label{mass2}
&&\left[1-2G_S\Pi_{\pi_+\pi_+}(k)\right]\left[1-2G_S\Pi_{\pi_-\pi_-}(k)\right] = 0\ , \nonumber\\
&&\left[1-2G_D\Pi_{D_1D_1}(k)\right]\left[1-2G_D\Pi_{D_{\bar 1}D_{\bar 1}}(k)\right] = 0\ ,\nonumber\\
&&\left[1-2G_D\Pi_{D_2D_2}(k)\right]\left[1-2G_D\Pi_{D_{\bar
2}D_{\bar 2}}(k)\right] = 0\ ,\nonumber\\
&&\det\left(\begin{array}{ccc}
1-2G_D\Pi_{D_3D_3}(k) &-2G_D\Pi_{D_3D_{\bar 3}}(k) \\
-2G_D\Pi_{D_{\bar 3}D_3}(k) &1-2G_D\Pi_{D_{\bar 3}D_{\bar 3}}(k) \\
\end{array}\right)\ =0\ .
\end{eqnarray}
For the meson and diquark masses computed through the above
dispersion relations at $k_0^2=M^2$ and ${\bf k}=0$, one needs to
know the explicit expressions $\Pi(k_0,{\bf 0})$ only. Performing
a relatively complicated but straightforward calculation including
Matsubara frequency summation, they can be written as
\begin{eqnarray}
\label{pi} &&\Pi_{\sigma\sigma}(k_0)= \Pi_{\pi_0\pi_0}(k_0) =
J_1(k_0^2)\ ,\nonumber\\
&&\Pi_{\pi_+\pi_+}(k_0)=\Pi_{\pi_-\pi_-}(-k_0)=J_2(\mu_e-k_0)\ ,\nonumber\\
&&\Pi_{D_1D_1}(k_0)=\Pi_{D_2D_2}(k_0)=\Pi_{D_{\bar 1}D_{\bar 1}}(-k_0)=\Pi_{D_{\bar 2}D_{\bar 2}}(-k_0)\nonumber\\
&&\ \ \ \ \ \ \ \ \ \ \ \ \ \ \ =I_\Delta+2(k_0+\mu_8)K_1(k_0)\ ,\nonumber\\
&&\Pi_{D_3D_3}(k_0)=\Pi_{D_{\bar 3}D_{\bar 3}}(-k_0)\nonumber\\
&&\ \ \ \ \ \ \ \ \ \ \ \ \ \ \
=I_\Delta+(4k_0^2-8\Delta^2)K_2(k_0^2)+8k_0K_3(k_0^2)\ ,
\nonumber\\
&&\Pi_{D_3D_{\bar 3}}(k_0)=\Pi_{D_{\bar
3}D_3}(k_0)=8\Delta^2K_2(k_0^2)\ ,
\end{eqnarray}
where the functions $K_1, K_2$ and $K_3$ related to the diquarks
are defined as
\begin{eqnarray}
&& K_1= \int{d^3{\bf p}\over (2\pi)^3}
\sum_{\epsilon=\pm}\Bigg[\left({1\over F_1^\epsilon}-{1\over
F_2^\epsilon}\right){f(E_{u3}^\epsilon)+f(E_{d3}^\epsilon)-1\over E^\epsilon_\Delta}\nonumber\\
&&\ \ \ \ \ +\left({1\over F_1^\epsilon}+{1\over
F_2^\epsilon}\right) {f(E_+^\epsilon)+f(E_-^\epsilon)-1\over
E_\Delta^\epsilon}\Bigg]\ ,\nonumber\\
&& K_2 = \int{d^3{\bf p}\over
(2\pi)^3}\sum_{\epsilon=\pm}\frac{1}{F_3^\epsilon}
\left(f(E_+^\epsilon)+f(E_-^\epsilon)-1\right)\ ,\nonumber\\
&& K_3 = -\int{d^3{\bf p}\over
(2\pi)^3}\sum_{\epsilon=\pm}\epsilon\frac{E_0^\epsilon}{F_3
^\epsilon}\left(f(E_+^\epsilon)+f(E_-^\epsilon)-1\right)\ ,
\end{eqnarray}
with
\begin{eqnarray}
F_1^\pm (k_0,|{\bf p}|)&=& k_0+\mu_8\mp E^\pm_0\mp E^\pm_\Delta \ ,\nonumber\\
F_2^\pm (k_0,|{\bf p}|)&=& k_0+\mu_8\mp E^\pm_0\pm E^\pm_\Delta \ ,\nonumber\\
F_3^\pm (k_0^2)&=&
E^\pm_\Delta\left(k_0^2-4\left(E^\pm_\Delta\right)^2\right)\ .
\end{eqnarray}

Since we will not discuss the meson masses in detail in this
Letter, the explicit expressions of the functions $J_1$ and $J_2$
related to the mesons are not listed here. However, the relations
among the meson dispersion relations shown in (\ref{pi}) can help
us to understand the meson mass splitting in chiral symmetry
restoration phase. From the first equation of (\ref{pi}) it is
clear that the masses of $\sigma$ and $\pi^0$ mesons become
degenerate in the color superconductivity phase,
\begin{equation}
M_\sigma=M_{\pi^0}\ .
\end{equation}
Remember that in presence of a nonzero electric chemical potential
$\mu_e$, the chiral symmetry $SU_L(2)\bigotimes SU_R(2)$ is
explicitly broken down to $U_L(1)\bigotimes U_R(1)$ with generator
$\tau_3$, and the masses of the three pion mesons are not the same
in the chiral restoration phase. This can be easily seen from the
difference among the three pion polarization functions shown in
(\ref{pi}). Only in the case of $\mu_e=0$, the relation
$J_2(-k_0)=J_1(k_0^2)$ results in degenerated meson mass $M_\sigma
=M_{\pi_0} =M_{\pi_+} =M_{\pi_-}$.

\begin{figure}
\centering \includegraphics[width=2.3in]{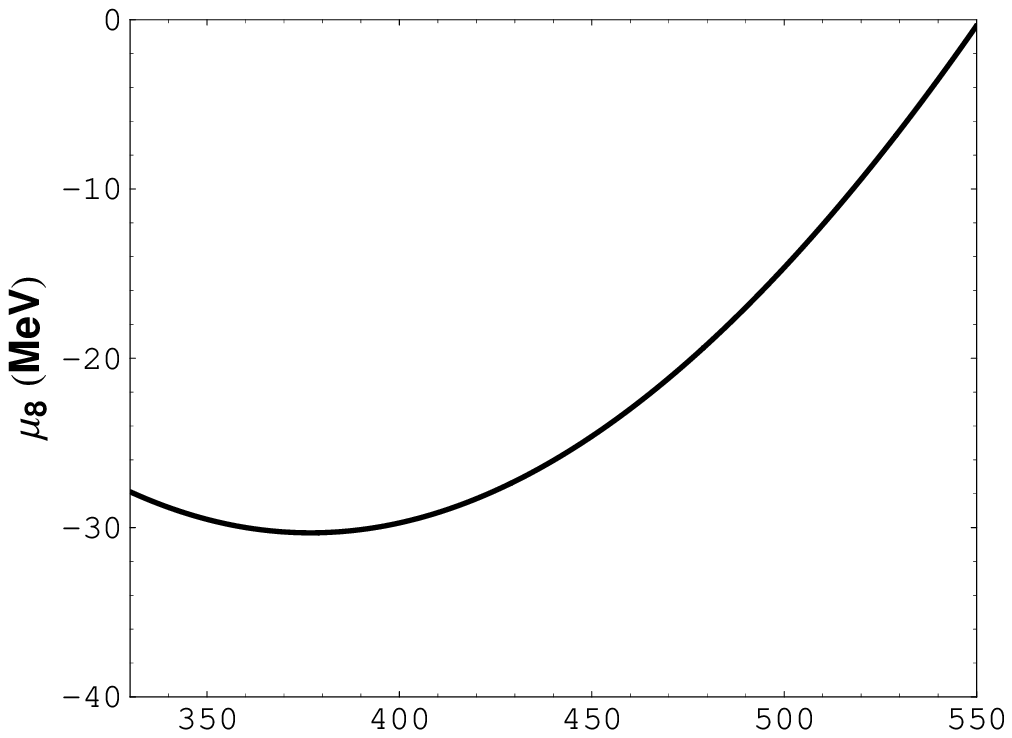}%
\hspace{0.5in}%
\includegraphics[width=2.3in]{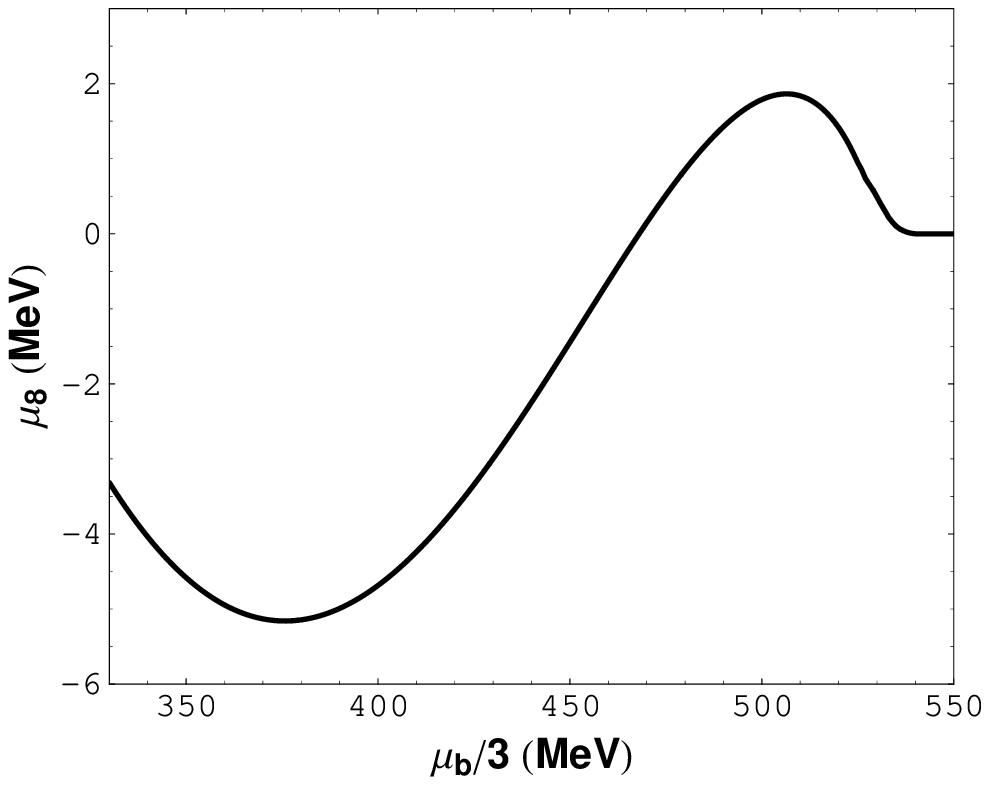}
\caption{The color chemical potential $\mu_8$ as a function of
baryon chemical potential $\mu_b$ at zero temperature in the color
superconductivity phase. The upper and lower panels are,
respectively, the results without ($\mu_e=0$) and with electrical
charge neutrality.} \label{fig1}
\end{figure}

Considering the gap equation (\ref{gap}) in the color
superconductivity phase and the relations among the polarization
functions $\Pi_{D_1D_1},\Pi_{D_2D_2},\Pi_{D_{\bar 1}D_{\bar 1}}$
and $\Pi_{D_{\bar 2}D_{\bar 2}}$ shown in (\ref{pi}), the two mass
equations for $D_1, D_2, D_{\bar 1}$ and $D_{\bar 2}$ are both
simplified as
\begin{equation}
(k_0^2-\mu_8^2)K_1(k_0)K_1(-k_0)=0\ .
\end{equation}
Obviously, one solution is $k_0^2=\mu_8^2$, and the other is
determined by
\begin{eqnarray}
H(k_0^2)=K_1(k_0)K_1(-k_0)=0.
\end{eqnarray}
From the comparison with the color charge density $Q_8$, one can
easily prove
\begin{eqnarray}
K_1(k_0=-\mu_8)=2{Q_8\over\Delta^2}\ .
\end{eqnarray}
Taking into account the color charge neutrality condition $Q_8=0$,
one has
\begin{equation}
H(k_0^2=\mu_8^2)=0.
\end{equation}
Therefore, the masses of the four diquarks, $D_1, D_2, D_{\bar 1}$
and $D_{\bar 2}$, are degenerate and exactly equal to the
magnitude of the color chemical potential,
\begin{equation}
M_{D_1}=M_{D_2}=M_{D_{\bar 1}}=M_{D_{\bar 2}}=|\mu_8|\ ,
\end{equation}
in the color superconductivity phase. The lower panel of
Fig.\ref{fig1} shows $\mu_8$ as a function of baryon chemical
potential $\mu_b$ at zero temperature, calculated through solving
the gap equation (\ref{gap}) and the color and electrical charge
neutrality condition (\ref{charge}). There are three parameters in
the NJL model (\ref{njl}) in chiral limit. The momentum cutoff
$\Lambda$ and the coupling constant $G_S$ can be fixed by fitting
the pion decay constant and chiral condensate in the
vacuum\cite{njlreview,hufner}, and the coupling constant $G_D$ in
the diquark channel is taken as $G_D=3G_S/4$\cite{neutral-2} in
our calculation. In this case the color superconductivity phase
starts at $\mu_b/3 = 330 MeV$. Since the above analytic discussion
for the meson and diquark masses does not depend on whether one
considers electrical charge neutrality or not, we show also in the
upper panel of Fig.\ref{fig1} the color chemical potential $\mu_8$
without considering electrical charge neutrality, namely we take
$\mu_e=0$ in the calculation. We see that in any case the
magnitude of $\mu_8$ is small. Especially, when we take the both
charge neutralities, $\mu_8$ is only a few MeV in a wide region.
That means, the four degenerated pseudo-Goldstone bosons are
almost massless in neutral color superconductor.

\begin{figure}
\centering \includegraphics[width=2.3in]{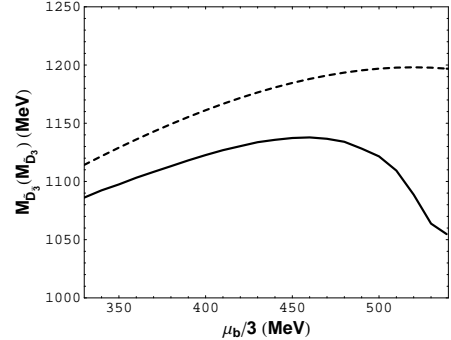} \caption{The
mass of the heavy diquark mode $\tilde D_{\bar 3} \left(\tilde
D_3\right)$, calculated through (\ref{heavy}), as a function of
baryon chemical potential $\mu_b$ at zero temperature in the color
superconductivity phase. The dashed and solid lines are,
respectively, the results without ($\mu_e=0$) and with electrical
charge neutrality.} \label{fig2}
\end{figure}

The relations among the polarization functions $\Pi_{D_3D_3},
\Pi_{D_3D_{\bar 3}}, \Pi_{D_{\bar 3}D_3}$ and $\Pi_{D_{\bar
3}D_{\bar 3}}$ and their explicit expressions shown in (\ref{pi})
reduce the mass equation for the modes $\tilde D_3$ and $\tilde
D_{\bar 3}$ which are linear combinations of $D_3$ and $D_{\bar
3}$ to
\begin{eqnarray}
k_0^2\left[(k_0^2-4\Delta^2)K_2^2(k_0^2)-4K_3^2(k_0^2)\right]=0\ .
\end{eqnarray}
One of its solution is of course
\begin{equation}
M_{\tilde D_3}\left(M_{\tilde D_{\bar 3}}\right) = 0\ ,
\end{equation}
and the other massive mode $\tilde D_{\bar 3} \left(\tilde
D_3\right)$ is calculated numerically through
\begin{eqnarray}
\label{heavy}
(k_0^2-4\Delta^2)K_2^2(k_0^2)-4K_3^2(k_0^2)=0\ .
\end{eqnarray}
Fig.\ref{fig2} shows the mass of this heavy mode as a function of
baryon chemical potential. It is around 1100 MeV and even more
heavy in the case without electric charge neutrality (dashed
line), in the color superconductivity phase.

We have investigated the quantum fluctuations in the neutral
two-flavor color superconductivity phase in mean field
approximation to quarks together with the random phase
approximation to mesons and diquarks in the frame of NJL model. We
have shown analytically that, there is only one massless Goldstone
boson, and the other four expected Goldstone bosons are actually
pseudo-Goldstone bosons, and their degenerated mass is exactly the
magnitude of the color chemical potential $\mu_8$ which is
introduced in the Lagrangian density to guarantee the color
neutrality of the system and breaks explicitly the color symmetry
from $SU_C(3)$ to $SU_C(2)\bigotimes U_C(1)$. By self-consistently
determining the diquark condensate, color and electric chemical
potentials, the pseudo-Goldstone mass is only a few MeV, the same
order like the current quark mass which breaks explicitly the
chiral symmetry of QCD. It is necessary to note that, while the
one Goldstone and four pseudo-Goldstone modes will finally be
eaten up by gauge fields, the nonzero color chemical potential
$\mu_8$ is also coupled to the meson and the heavy diquark modes
which will then be reflected in the measurable properties of the
system.

{\bf Acknowledgement:} We thank M.Alford, M.Huang and I.Shovkovy
for helpful comments. The work is partially supported by the
Natural Science Foundation of China.

\end{document}